\documentclass[twocolumn]{aastex63}

\def\hii{\mbox{H\,{\sc ii}}}
\newcommand*{\textquotedouble}[1]{\textquotedblleft #1\textquotedblright}
\usepackage{scalerel}
\usepackage{rotating}

\shorttitle{Sample article}
\shortauthors{Veena et al.}

\begin{document}

\title{A Kpc Scale Molecular Wave in the Inner Galaxy: Feather of the Milky Way?}

\correspondingauthor{V. S Veena}
\email{veena@ph1.uni-koeln.de}

\author{V. S Veena}
\affiliation{I. Physikalisches Institut, Universit\"at zu K\"oln, Z\"ulpicher Str.77, 50937, K\"oln, Germany}

\author{P. Schilke}
\affiliation{I. Physikalisches Institut, Universit\"at zu K\"oln, Z\"ulpicher Str.77, 50937, K\"oln, Germany}

\author{\'A. S\'anchez-Monge}
\affiliation{I. Physikalisches Institut, Universit\"at zu K\"oln, Z\"ulpicher Str.77, 50937, K\"oln, Germany}

\author{M.  C.  Sormani}
\affiliation{Institut f\"ur Theoretische Astrophysik, Zentrum f\"ur Astronomie, Universit\"at Heidelberg, Albert-Ueberle-Str 2, D-69120, Heidelberg, Germany}

\author{R.  S. Klessen}
\affiliation{Institut f\"ur Theoretische Astrophysik, Zentrum f\"ur Astronomie, Universit\"at Heidelberg, Albert-Ueberle-Str 2, D-69120, Heidelberg, Germany}

\author{F.  Schuller}
\affiliation{Leibniz-Institut f\"ur Astrophysik Potsdam (AIP), An der Sternwarte 16, 14482, Potsdam, Germany}

\author{D. Colombo}
\affiliation{Max-Planck-Institut f\"ur Radioastronomie, auf dem H\"ugel 69, 53121 Bonn, Germany}

\author{T. Csengeri}
\affiliation{Laboratoire d'Astrophysique de Bordeaux,  Univ. Bordeaux, CNRS, B18N, all\'e\'e Geoffroy Saint-Hilaire, 33615, Pessac, France}

\author{M. Mattern}
\affiliation{Laboratoire d'Astrophysique (AIM), CEA, CNRS, Universit\'e Paris-Saclay,  Universit\'e Paris Diderot, Sorbonne Paris Cit\'e, 91191 Gif-sur-Yvette, France}

\author{J.  S. Urquhart}
\affiliation{Centre for Astrophysics and Planetary Science, University of Kent, Canterbury CT2 7NH, UK}

\begin{abstract}
We report the discovery of a velocity coherent,  kpc-scale molecular structure towards the Galactic center region with an angular extent of 30$^\circ$ and an aspect ratio of 60:1.  The kinematic distance of the CO structure ranges between 4.4 to 6.5 kpc.  Analysis of the velocity data and comparison with the existing spiral arm models support that a major portion of this structure is either a sub-branch of the Norma arm or an inter-arm giant molecular filament, likely to be a kpc-scale feather (or spur) of the Milky Way,  similar to those observed in nearby spiral galaxies.  The filamentary cloud is at least 2.0 kpc in extent, considering the uncertainties in the kinematic distances, and it could be as long as 4 kpc. The vertical distribution of this highly elongated structure reveals a pattern similar to that of a sinusoidal wave. The exact mechanisms responsible for the origin of such a kpc-scale filament and its wavy morphology remains unclear.  The distinct wave-like shape and its peculiar orientation makes this cloud,  named as the Gangotri wave,  one of the largest and most intriguing structures identified in the Milky Way. 
\end{abstract}


\section{Introduction} \label{sec:intro}

Characterising the large-scale spiral structure of the Milky Way remains a challenge since the solar system lies within the Galaxy.  Most Milky Way models suggest the presence of a central bar besides four major gaseous spiral arms and some extra arm segments and spurs \citep{2019ApJ...885..131R}.  This is derived from the velocity data obtained from the distribution of atomic,  molecular and ionized hydrogen,  CO molecule as well as parallax information from stellar sources within the Milky Way \citep[e.g.,][]{{1958MNRAS.118..379O},{1976A&A....49...57G},{2001ApJ...547..792D}, {2018A&A...616A..11G}}.  The observational studies of external spiral galaxies reveal that spiral arms are not smooth and continuous features.  Instead,  they mostly comprise multiple sub-structures that are referred to as spurs,  branches or feathers,  and which give the arms their clumpy or patchy appearance \citep{{1970IAUS...39...22W},{1970IAUS...38...26L},{1985IAUS..106..255E},{2006ApJ...650..818L},{2017ApJ...836...62S}}.

\par With the advent of sensitive and high-resolution infrared and millimeter surveys,  there has been tremendous progress in unveiling the intricate morphology of the large-scale gas structures in our Galaxy.  For example, the identification of multiple giant molecular filaments (GMFs) on scales of hundreds of parsecs has redefined our view of the Milky Way's structural dynamics \citep[e.g.,][]{{2010ApJ...719L.185J},{2014A&A...568A..73R},{2015MNRAS.450.4043W},{2016A&A...590A.131A},{2018ApJ...864..153Z}}.  Few of these extremely long and narrow GMFs appear to lie parallel to and within few parsecs from the Galactic mid-plane.  Their line-of-sight velocities suggest that GMFs are often associated with nearby spiral arms.  They are frequently called \textquotedouble{bones/spines},  and trace the densest regions of the Galactic spiral arms.  Hence,  they could potentially be used to map the skeleton of our Galaxy \citep[e.g.,][]{2014ApJ...797...53G}.  In an alternate scenario, they could be \textquotedouble{spurs/feathers} emanating from spiral arms or inter-arm clouds similar to that in simulations or observed in nearby galaxies \citep[e.g.][]{{2014ApJ...784....3C},{2014A&A...568A..73R},{2017MNRAS.470.4261D},{2020MNRAS.492.1594S}}. 

\par In this work,  we report the discovery of a highly unusual filamentary cloud located within 5~kpc of the Galactic center region ($325^\circ<l<355^\circ$).  The extreme length,  sinusoidal wave-like morphology and relative orientation with respect to the adjacent Norma and 3-kpc spiral arms make this structure one of the most unusual features identified in the Milky Way.  Given the uniqueness of this molecular structure,  we name it as \textquotedouble{Gangotri wave}, after the Gangotri glacier,  the primary source of the Ganges, which is the longest river in India.  The observations, data analysis and results are presented in Sections 2 and 3 respectively and discussed in detail in Section 4.  Finally,  we present our conclusions in Section 5.


\section{Data} \label{sec:data}

\subsection{SEDIGISM}

The SEDIGISM (Structure, Excitation, and Dynamics of the Inner Galactic Interstellar Medium) project is a CO survey covering 78~deg$^2$ of the inner Galaxy ($-60$~deg$<l<+18$~deg, $\mid b\mid<0.5$~deg) in the $J=2-1$ rotational transition of $^{13}$CO and C$^{18}$O \citep{{2017A&A...601A.124S},{2021MNRAS.500.3064S},{2021MNRAS.500.3027D},{2021MNRAS.500.3050U}}.  The $^{13}$CO isotopologue of CO is less abundant than $^{12}$CO by factors up to 100,  making it an ideal tracer of the cold and dense molecular clouds of the Milky Way.  The entire 78~deg$^2$ survey area is divided into $2.0\times1.0$~deg$^{2}$ fields.  The spatial and spectral resolutions are 28$\arcsec$ and 0.25~km\,s$^{-1}$ respectively,  and the typical r.m.s noise in each field is $\sim$0.8~K.  In order to search for potential giant filaments in the inner Galaxy,   we have combined all the individual tiles to produce 78 deg$^2$ mosaic maps of $^{13}$CO and C$^{18}$O emission.  For this,  we have first smoothed the individual tiles to a lower velocity resolution of 1.5~km\,s$^{-1}$.  We then combined the smoothed tiles to create a single mosaic using the Montage software \citep{2003ASPC..295..343B}.  In this paper, we have used a spatially smoothed mosaic with an angular resolution of 5$\arcmin$ for a better visualisation of the features.

\subsection{ThrUMMS}
The Three-mm Ultimate Mopra Milky Way Survey \citep{2015ApJ...812....6B} is the millimeter wave molecular-line mapping survey that covers $60^\circ\times2^\circ$ of the fourth quadrant of the Galaxy in the $J=1-0$ rotational transition of $^{12}$CO, $^{13}$CO,  C$^{18}$O and CN at spatial and spectral resolutions of 1$\arcmin$ and 0.3 km/s, respectively. We have used the  $^{12}$CO data tiles to create a mosaic for our analysis.

\subsection{SGPS}
The Southern Galactic Plane Survey (SGPS) \citep{2005ApJS..158..178M} is a H{\scshape i} survey combining data from the Australia Telescope Compact Array and the Parkes Radio Telescope covering 325 deg$^2$ at arcminute resolution. We have used the SGPS I data tiles with an angular resolution of $2.2\arcmin$ and spectral resolution of 0.8~km/s.

\begin{sidewaysfigure*}
\begin{minipage}[t]{.3\textwidth}
\centering
\vspace*{8 cm}
\hspace*{-0.2 cm}
\includegraphics[scale=1.48]{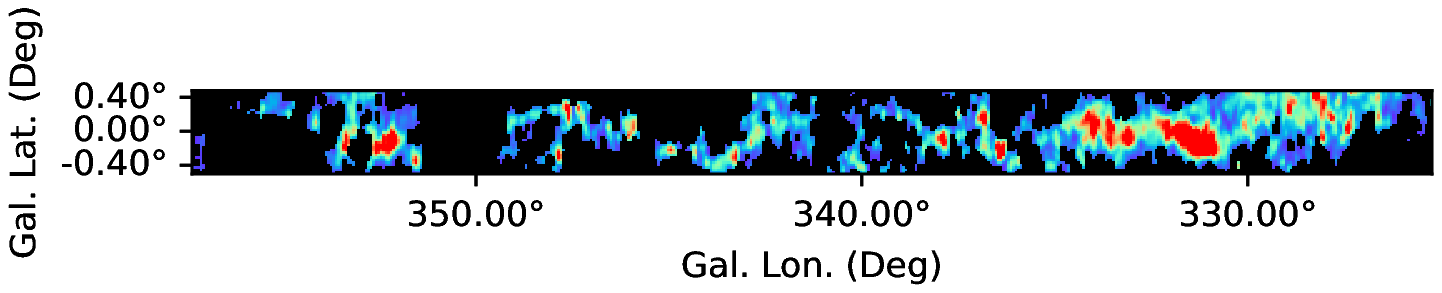} \quad \includegraphics[scale=1.45]{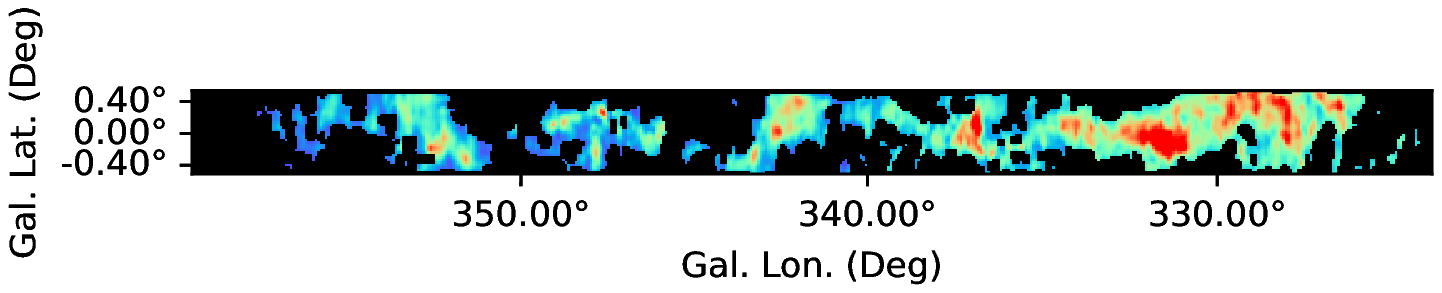} 
\end{minipage}
\caption{(top) $^{13}$CO integrated intensity map from SEDIGISM survey in the velocity range $-95$ to $-75$~km/s showing the wave-like feature.  (bottom) $^{12}$CO integrated intensity map from ThrUMMS survey in the same velocity range as the top panel, smoothed to an angular resolution of 5$\arcmin$.  Images are stretched along the Y axis for a better visualization.}
\label{wavem0}
\centering
\includegraphics[scale=1]{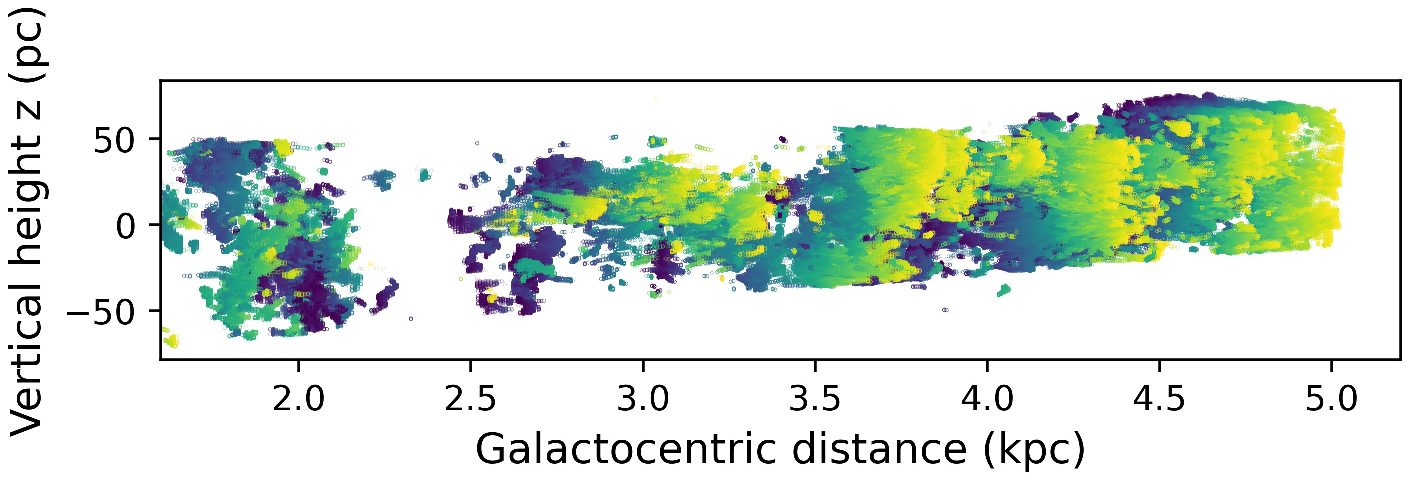} 
\caption{Galactocentric distance versus vertical distance from the Galactic plane (z) of the cloud in the velocity range [$-95,-75$]~km/s showing the vertical oscillations with respect to the Galactic mid-plane. }
\label{lon_z}
\end{sidewaysfigure*}

\section{Results} \label{sec:result}

In order to search for potential bones/spurs in the CO emission, we have used the $^{13}$CO mosaic from SEDIGISM survey.  Integrated intensity maps were generated at velocity intervals of 10~km/s. The resultant maps were visually inspected to identify filamentary features.  Fig.~\ref{wavem0}(top) shows the $^{13}$CO integrated intensity map of the region from $325^\circ < l< 358^\circ$ in the velocity range $-95$~km/s to $-75$~km/s.  The map reveals an emission feature that is extending up to 30$^\circ$ in longitude from $325^\circ <l <355^\circ$.  The ratio of angular length (30$^\circ$) to angular width ($\sim0.5^\circ$) implies a projected aspect ratio of 60:1.  This illustrates the highly elongated/filamentary morphology of the feature,  which is also evident in the $^{12}$CO integrated map (Fig.~\ref{wavem0}(bottom)) obtained from ThrUMMS.  The SEDIGISM $^{13}$CO channel maps toward the region are presented in the Appendix. 

\subsection{Distance and size of the filament}\label{sec:kdist}

There are no previous distance estimates of the filament available in the literature.  As there are no trigonometric parallaxes,  it is difficult to obtain accurate values.  However, an approximate distance can be obtained using the kinematic method \citep[e.g.][]{2009ApJ...699.1153R}.  For this,  the velocity distribution of the $^{13}$CO emission can be used.  In order to obtain the velocity distribution towards the filament,  we availed the  fully automated Gaussian decomposer \texttt{GAUSSPY+} \citep{{2015AJ....149..138L},{2019A&A...628A..78R}}.  It makes use of the \texttt{GAUSSPY} algorithm to analyze the complex spectra extracted from the Galactic plane by decomposing them into multiple Gaussian components.  For the spectral decomposition,  we subdivided the 30$^\circ$ mosaic into sub-mosaics of size 5$^\circ$ each.  We have taken the default parameters for the decomposition provided by \citet{2019A&A...628A..78R},  except for the smoothing parameters $\alpha_1$ and $\alpha_2$.  To determine these parameters, we generated the training sets that were used to train \texttt{GAUSSPY+} with 1000 randomly selected spectra from CO sub-cubes with sizes 5$^\circ$ each.  The decomposition of the entire 30$^\circ$ mosaic resulted in $13.2\times10^6$ Gaussian components.  Since the filamentary structure is mainly identified in the velocity range [$-95,-75$] km/s,  we excluded all the components outside this specified velocity range from further analysis.

\par As the filament is located inside the solar circle,  the distance calculations are hampered by the kinematic distance ambiguity (KDA).  For each radial velocity along a given line of sight,  there are two possible values,  one corresponding to the near distance and the other to the far distance.  In order to resolve the KDA,  the Hi-GAL compact source catalogue \citep{2017MNRAS.471..100E} is used.  Only sources with LSR velocity matching that of the wavy filament and having a reliable distance estimate are selected for the analysis.  According to \citet{2017MNRAS.471..100E},  these reliable values are determined by first estimating the kinematic distance to the sources using the CO data and the Galactic rotation curve model of \citet{1993A&A...275...67B}.  Then,  the KDA is resolved by matching positions with a catalogue of sources with known distances (such as \hii~regions and masers),  or alternatively,  with features in extinction maps.  The distribution of Hi-GAL sources correlate well with that of the filament (see Fig.~\ref{wavem0}).  The distances of the Hi-GAL sources fall in the range $4.5-7.0$~kpc,  favoring the near distance for the filament.  In addition to this,  the infrared star count and extinction map towards the fourth quadrant of the Galaxy \citep{2019MNRAS.488.2650S} are also utilized to resolve the KDA.  These maps show that there is relatively lower stellar density and higher dust extinction towards the region corresponding to the position of the filament.  This again implies that the filament is likely to be located at the near side.

\par The heliocentric distance to the filament at $l\sim325^\circ$ is $\sim4.5$~kpc,  whereas it extends farther away,  towards the Galactic center region (distance$>$7~kpc) at higher longitudes.  Since there are large uncertainties involved in the kinematic distance estimates beyond $l\sim350^\circ$, getting accurate distance distribution of the filament is highly challenging.  Using the derived kinematic distances, the de-projected length of the GMF is estimated to be 4~kpc. Assuming that the distance to the filament is 4.4~kpc,  we get a lower limit to the spatial extent of the filament of 2.4~kpc.  The average vertical width of the filament is under 50~pc,  resulting in aspect ratio of 60:1.

\begin{figure*}
\centering
\hspace*{-1.5 cm}
\includegraphics[scale=0.42]{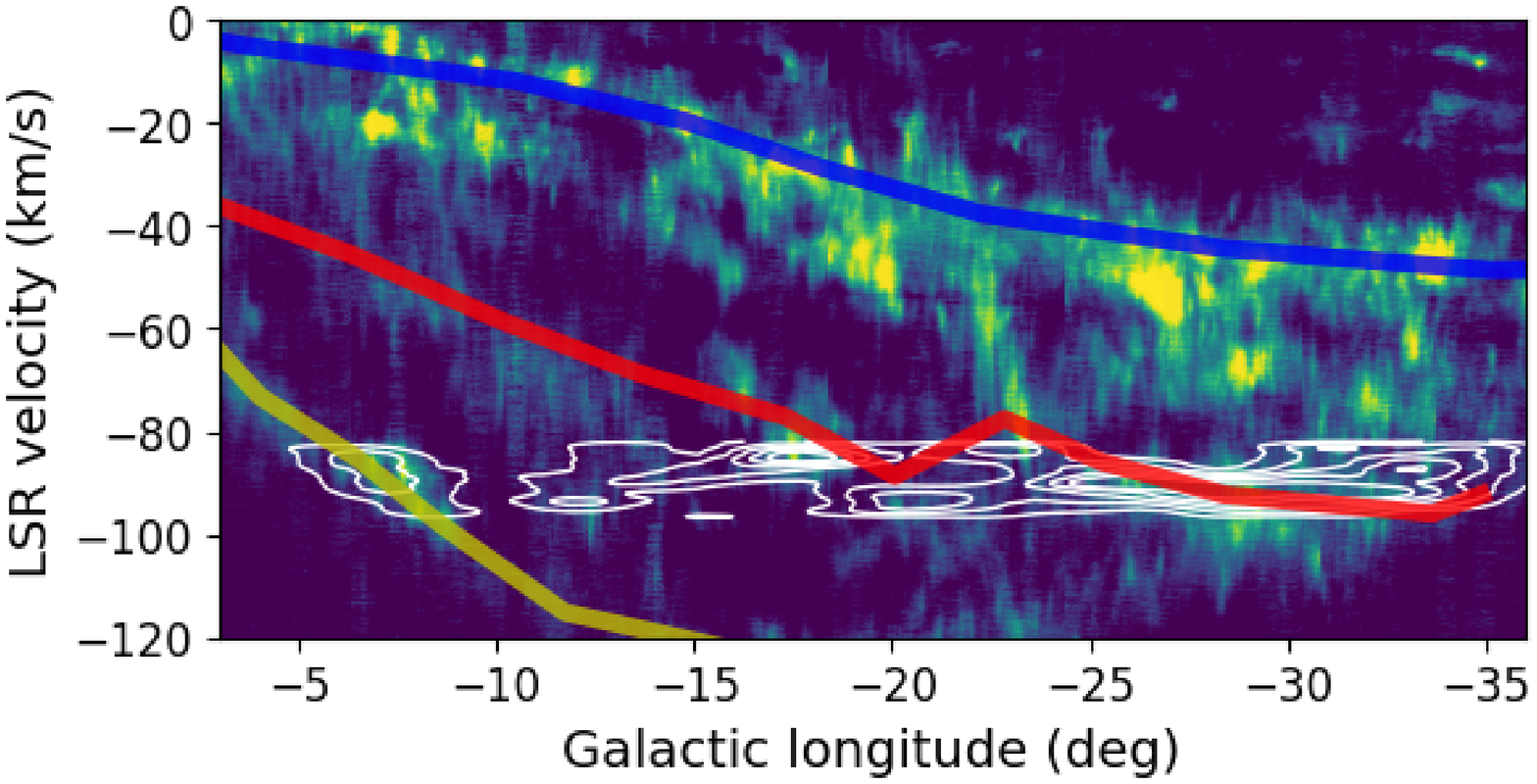} \quad\includegraphics[scale=0.24]{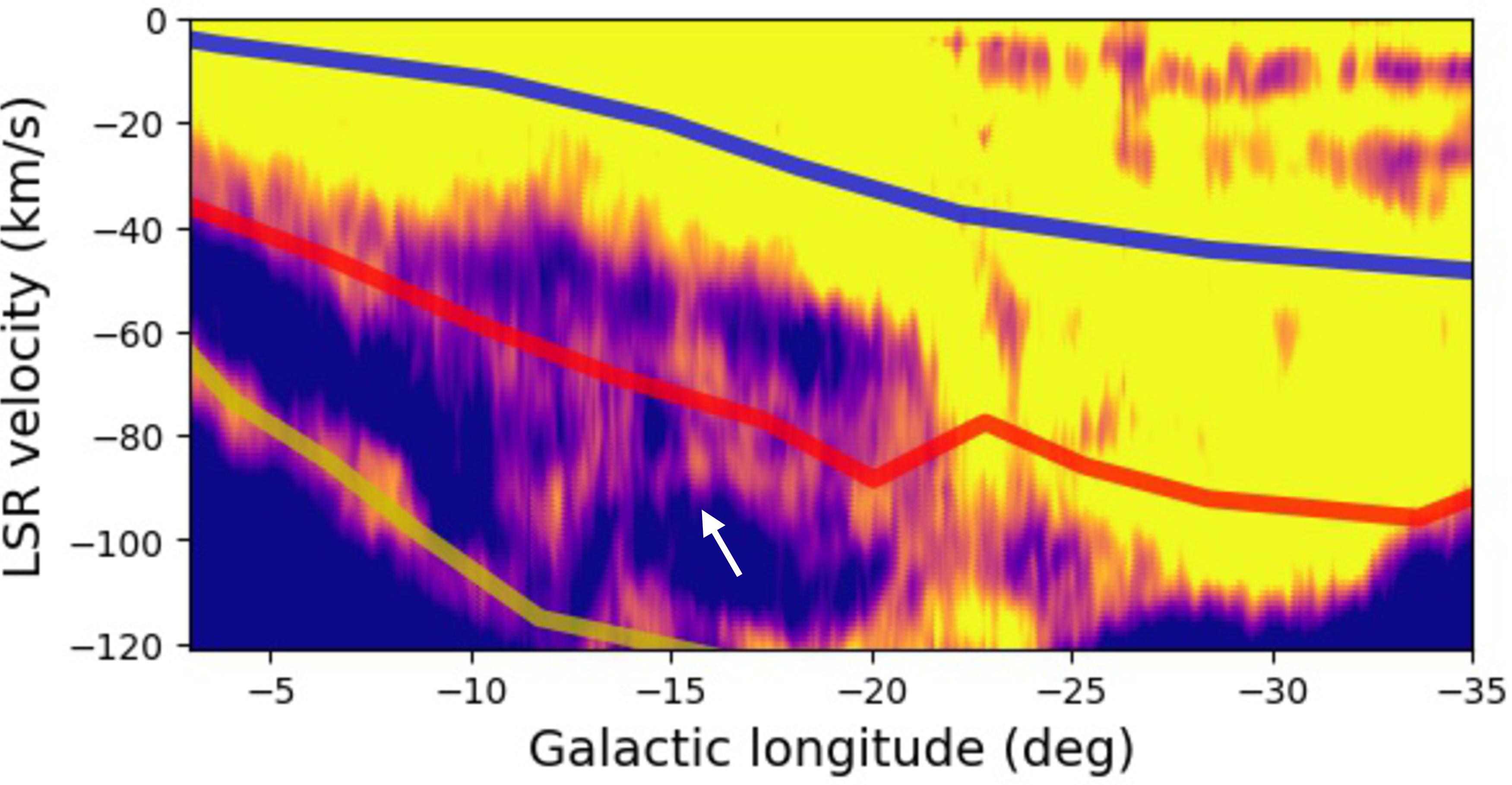}
\caption{(Left) $^{12}$CO emission from ThrUMMS as a function of LSR velocity and Galactic longitude.  The coloured lines (blue: Scutum-Centaurus arm, red: Norma arm, yellow: 3-kpc arm) corresponds to the spiral arm models of \citet{2016ApJ...823...77R}. Overplotted are the $^{12}$CO contours corresponding to Gangotri wave. (Right) H{\scshape i} emission from SGPS survey as a function of LSR velocity and Galactic longitude overlaid with the spiral arm models of \citet{2016ApJ...823...77R}.  The Gangotri wave is seen as a distinct structure branching out from the main Norma spiral arm emission, pointed with an arrow. }
\label{PV}
\end{figure*}

\subsection{Morphology and the three dimensional orientation of the filament} 

Apart from its velocity coherent distribution at kpc-scales,  the GMF also exhibits an intriguing two-dimensional projected morphology.  The visual inspection of the integrated intensity map shows a sinusoidal wave-like pattern with corrugations in the Galactic latitude along the GMF.  The undulations appear to be dampening towards the right hand side ($l<335^\circ$) of the GMF similar to that of the recently discovered Radcliffe wave \citep[][see Section 4.2.2]{2020Natur.578..237A}.  Fig.~\ref{lon_z} shows the distribution of the GMF with respect to the plane of the Milky Way.  The vertical oscillations span $\sim\pm70$~pc in the Galactic mid-plane.  These oscillations are within the average thickness of the molecular disc of the Milky Way \citep[$\sim160$ pc; e.g.,][]{2006PASJ...58..847N}.  Assuming the lower limit distance of 4.5~kpc,  the wavelength (towards the left side of the GMF, i.e., $l\sim350^\circ$) is estimated to be 600~pc.  There appears to be a break in the wave at $l\sim350^\circ$ in the $^{13}$CO map that is less prominent in the $^{12}$CO intensity map that could be due to lack of high density molecular gas in the region, possibly from star formation activity in the past.  The wave is relatively narrower towards the East (i.e., $l>345^\circ$).  Towards the West,  where the velocities are consistent with the Norma arm,  the emission is more diffuse in nature.  This could also be due to the foreground or background contamination from other molecular clouds along the Norma arm.  There are numerous \hii~regions,  dense cores and infrared bubbles \citep[e.g., MWP1G352880-002000S,][]{2012MNRAS.424.2442S} located along the filament indicating the star formation flurry of the cloud. The wavy structure is not evident in the submillimeter continuum map (see Appendix) due to possible contamination from molecular clouds in the foreground and background.

\subsection{Column density and mass}

The total column density of the wavy filament is estimated using the standard method described in earlier studies \citep[e.g.,][]{2013A&A...560A..24C}.  In order to estimate the column density of the filament,  we have converted the integrated intensity of $^{13}$CO(2--1) emission into H$_2$ column density using a mean conversion factor,  $X_{{^{13}_\textrm{\scaleto{CO(2-1)}{8pt}}}}$ = $1.08\times 10^{21}$~cm$^{-2}$~(K km/s)$^{-1}$ adopted from \citet{2017A&A...601A.124S}.  A column density map of the filament has been obtained.  From the map, the peak and mean column densities along the filament are estimated as $2.1\times10^{23}$~cm$^{-2}$,  and $6.1\times10^{21}$~cm$^{-2}$, respectively. 

The total column density of the filament can be converted directly to the mass of the filament using the expression

\begin{equation}
M=N(\textrm{H}_2) \mu m_\textrm{H} A.
\end{equation}

\noindent Here $M$ represents the mass of the filament,  $A$ is the area of the filament,  $\mu$ is the mean weight of the molecular gas taken to be 2.86 assuming that the gas is 70$\%$ molecular hydrogen by mass \citep{2010A&A...518L..92W} and $m_\textrm{H}$ is the mass of the hydrogen atom.  Using this equation,  we find the total mass of the cloud as $8.7\times10^6$~M$_\odot$.  We point out that this is a lower limit as we estimated the mass assuming a lower limit distance to the filament,  that is 4.5~kpc.

\subsection{Velocity structure of the filament and location in the Galaxy}\label{sec:lv}

The Galactic longitude versus LSR velocity ($l$,V) diagram can be used to understand the velocity structure of the wavy filament and its approximate location within the Milky Way.  Fig.~\ref{PV}(Left) shows the distribution of the wavy filament over the Galactic CO emission. Overplotted are the spiral arm models of \citet{2016ApJ...823...77R}.  From the figure, it is evident that in the longitude range below 343$^\circ$, the emission matches well with the model of Norma spiral arm.  In the longitude range 343$^\circ$-350$^\circ$,  the velocity of the structure deviates from the Norma arm with maximum deviated velocity of $\sim35$km/s.  Beyond 350$^\circ$ longitude, the emission matches quite well with that of the 3-kpc spiral arm model.  The break in the wavy structure observed in the integration emission maps is also seen in the ($l$,V) diagram.  A possible explanation is that the wavy filament is not a coherent structure and rather a superposition of features from Norma and 3-kpc arms.  In order to investigate this further, we have also generated H{\scshape i} ($l$,V) diagram using SGPS data.  Since H{\scshape i} traces low density gas compared to CO,  any connection should be prominent in the H{\scshape i} plot. Fig.~\ref{PV}(Right) shows the H{\scshape i} ($l$,V) diagram.  In the plot, the emission corresponding to the wavy structure appears very prominent in the Norma-3-kpc inter-arm region.  The structure appears to be branching out from the main Norma arm at longitudes beyond 340$^\circ$ extending further towards the direction of 3-kpc arm.  This implies that the structure is not entirely confined to the Norma arm.  The structure is  2 kpc in extent,  excluding emission from the 3 kpc region.  Even though the emission corresponding to the 3-kpc arm at present appears to be disconnected from the structure connected to the Norma arm,  the vertical oscillation seems to transcend this gap. Either the connection existed in the past, or the cause for the vertical oscillation does not require a physical connection of the gaseous structures, which would be the case for example,  waves in the gravitational potential. The velocity distribution of the entire structure also shows fluctuations along the Galactic longitude, evident in the ($l$,V) diagrams. Such large-scale velocity fluctuations and oscillatory gas flows are detected in molecular clouds toward the Galactic and extragalactic environments that may form via gravitational instabilities \citep{2020NatAs...4.1064H}.

\section{Discussion} \label{sec:discussion}

\subsection{Orientation of the GMF with respect to the spiral arms}

\begin{figure}
\centering
\hspace*{-1.0 cm}
\includegraphics[scale=0.45]{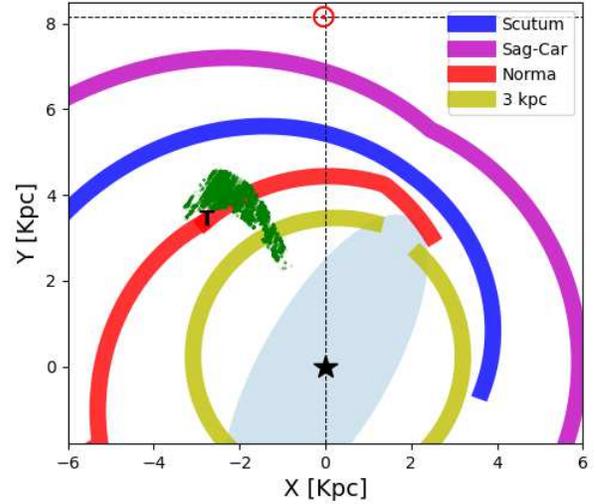} 
\caption{Distribution of the GMF with respect to the spiral arm models of \citet{2019ApJ...885..131R}.  The location of the Norma arm tangent point is also marked.}
\label{reid_wave}
\end{figure}

In order to better understand the physical origin of this molecular wavy structure,  an investigation on its large-scale distribution with respect to the spiral arms of the Galaxy is required.  The most important question is whether the cloud belongs to a spiral arm or it is an inter-arm spur.  It is seen from Section~\ref{sec:lv} that the ($l$,V) distribution of the wavy structure is not consistent with one particular spiral arm.  Using the kinematic distance estimates in Section~\ref{sec:lv} and the corresponding Galactic longitudes combined with the spiral arm models of \citet{2019ApJ...885..131R},  a planar view of the distribution of wavy filament with respect to the spiral arms of the Milky Way can be determined,  as indicated in Fig.~\ref{reid_wave}.  The major fraction of the structure on the near side appears to be within the Norma arm. However, it extends farther out into the inter-arm region.  This means that the filament being a bone/spine can be ruled out,  as such features, per definition,  are closely associated with the spiral arms \citep{2014ApJ...797...53G}.  At the distance range corresponding to the Gangotri wave, the typical spiral arm width is estimated around $100-250$pc \citep{2019ApJ...885..131R}.  If we exclude the emission matching with Norma and 3-kpc arms, the filament is extending over 500 pc in the inter-arm region (assuming a minimum heliocentric distance of 4.5 kpc).  \citet{2021A&A...651L..10K} recently identified a 1 kpc, high pitch angle and high aspect ratio (7:1) structure associated with the Sagittarius arm. This feature is believed to be either an isolated entity, a sub-structure within the spiral arm or an inter-arm spur.  In case of the Gangotri wave, although the connection with 3-kpc arm is not firm, the observed orientation suggests that the structure is either a sub-arm/branch of the Norma arm or an inter-arm spur, similar to those observed in external spiral galaxies.


\par
\begin{figure}
\centering
\includegraphics[scale=0.66]{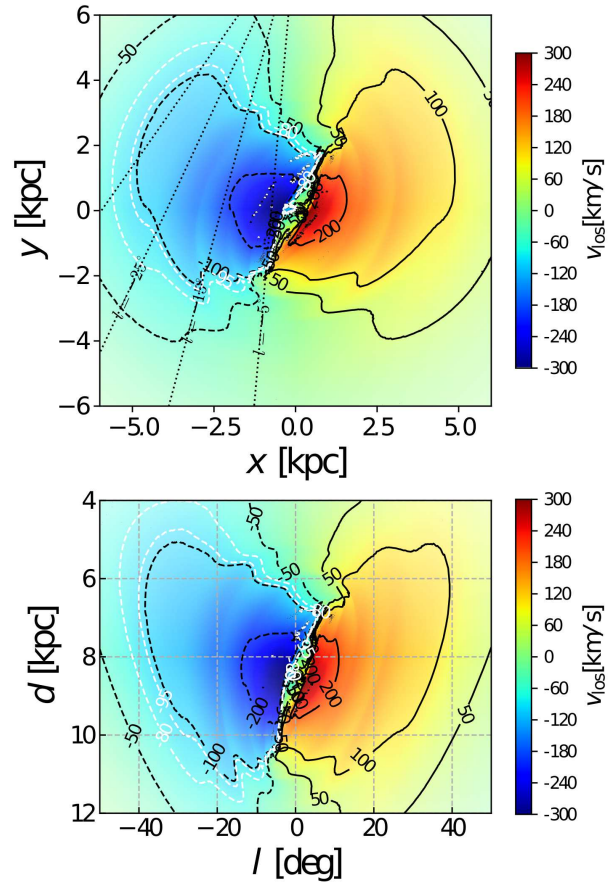} 
\caption{Mass-averaged line-of-sight velocity in the (x,y) plane (top) and (l,d) plot (bottom) for a snapshot of the simulations in \citet{2018MNRAS.475.2383S}.  The snapshot is similar to those shown in Figures 8, 9 and 10 in the paper.  The labelled contours indicate the line-of-sight velocities.  The white-dashed contours represent velocities of $-80$ and $-95$ km/s,  corresponding to that of Gangotri  wave, according to the model.  Here,  $l$ and d corresponds to Galactic longitude and distance from the Sun,  and (x, y) corresponds to Galactocentric coordinates respectively.}
\label{wave_nonrot}
\end{figure}

 We have also used the numerical models to analyse the impact of non-circular motions induced by the bar near the Galactic center on the orientation and distance of Gangotri wave \citep[e.g.,][]{1999A&A...345..787F}.  Fig.~\ref{wave_nonrot} shows the orientation and distance of the  based on the hydrodynamical simulations of \citet{2018MNRAS.475.2383S}.  The model includes a rigidly-rotating barred spiral potential fitted to the Milky Way and allows us to account for the errors in kinematic distances due to the non-circular motions driven by the bar.  From the plots,  it is evident that the kinematic distance of the Gangotri wave,  after correcting for the non-circular motions lie in the range $\sim4.0-6.5$~kpc.  Thus,  this confirms the sub-arm/inter-arm nature of the wave.

\par \citet{1970IAUS...39...22W} noted that many of the nearby spiral galaxies have spurs,  that appear to originate on the outside of a spiral arm and extend into the inter-arm regions.  Strong dust lanes along the inner edge of the arms identified in some of the spiral galaxies are often termed as feathers \citep{{1970IAUS...38...26L},{1973ApJ...179..755P}}.  An HST survey by \citet{2006ApJ...650..818L} discovered feathers toward 45 spiral galaxies.  They have concluded that the feathers are common in prototypical Sc galaxies as well as in Sb galaxies.  The typical feathers are described as dark,  delineated extinction features,  that emerge from primary dust lanes (PDL) that run along the inside part of the arm and then extend well into the inter-arm regions.  Some of these feathers are several kpc in extent and appear to merge with the PDL of the next arm (e.g., Fig.~1 and 2 of \citealt{2006ApJ...650..818L}).  As the Milky way is a barred spiral galaxy,  we expect it to have such inter-arm feathers as well.  Thus,  the newly identified Gangotri wave could be a potential feather of the Milky Way.

\subsection{Physical origin of the Gangotri wave}

\subsubsection{Inter-arm nature: Galactic feather?}

From the earlier section,  it is clear that Gangotri wave is a kpc-scale Norma sub-arm or inter-arm spur possibly connecting the near side of the Norma and 3-kpc spiral arms.  The formation of such spiral arm sub-structures,  such as feathers and spurs in global magnetohydrodynamic (MHD) simulations has been explored in detail in many works \citep[e.g.,][]{{2002ApJ...570..132K},{2006ApJ...647..997S}}. There is still no consensus on the mechanism that generates these substructures.  Proposals include self-gravity,  magneto-Jeans instabilities,  the wiggle instability,  shear due to differential rotation,  the effect of correlated SN feedbacks. 

The self-gravity and magnetic fields within the spiral arms lead to the rapid growth of overdensities.  The differential compression of gas flowing through such arms could then result in the formation of sheared structures in the inter-arm regions.  Simulations of GMFs in spiral galaxies have furthermore demonstrated that GMFs can be formed from gas clouds that exit a spiral arm and get stretched by the differential rotation of the gas \citep{{2014MNRAS.441.1628S},{2017MNRAS.470.4261D}}.  These GMFs can span lengths $>500$~pc in H$_2$.  However,  CO is seen to be present only on the denser regions of the clouds,  putting an upper limit on length of CO filaments as $\sim$100 pc. 

\par The wiggle instability (WI) of spiral shocks in a galactic disk has been proposed as the mechanism responsible for the formation of gaseous feathers in spiral galaxies.  According to  \citet{2004MNRAS.349..270W},  the WI is a manifestation of Kelvin-Helmholtz (KH) instability of shear layer behind the shock.  This view has been challenged by later works,  where the WI is proven to be distinct from the KH instability \citep{{2014ApJ...789...68K},{2017MNRAS.471.2932S}}.  3D ISM simulations including the effects of star formation and supernova (SN) feedbacks (TIGRESS) by \citet{2020ApJ...898...35K} have shown that the correlated SN feedback produces short lived spurs/feathers ($t\sim30$~Myr) with magnetic fields parallel to their length,  in contrast to long-lived features ($t\sim200$~Myr) with perpendicular magnetic fields induced by gravitational instabilities.  Magnetic field observations of the spur/feather could therefore be used to find vital clues regarding whether the structure is produced by magneto-Jeans instability or SN feedbacks.  It is possible that the feather-like nature of the Gangotri wave is due to a combination of all of the above mechanisms.

\subsubsection{Wavy morphology}

Apart from its kpc-scale extent and the inter-arm nature,  another intriguing feature of Gangotri wave is its sinusoidal wave-like morphology.  Small-scale bending waves are detected in the dust lanes of five nearby edge-on galaxies \citep{2020MNRAS.495.3705N}.  The corrugation amplitude is found in the range of $70-200$~pc with the corresponding wavelengths in the range $1-5$~kpc.  These corrugations are seen in stars,  gas and dust,  indicating that they are most likely to be caused by gravitational instabilities. 

\par A sine wave-like structure perpendicular to the plane is observed in the vertical profile of the Carina-Sagittarius spiral arm ($-70^\circ\leq l\leq 30^\circ$) obtained from the young open cluster distribution \citep{1992ApJ...399..576A}.  The wavelength of the structure is $\sim2.4$~kpc,  which is similar to the distribution of atomic hydrogen in the same spiral segment.  Three-dimensional density waves are envisaged as the likely mechanism responsible for the observed vertical structure.  \citet{2020Natur.578..237A} identified a narrow and coherent 2.7~kpc structure in the solar neighborhood.  The structure,  known as the Radcliffe wave,  contains many of the clouds thought to be associated with the Gould belt.  The 3D distribution of the Radcliffe wave is described by a damped sinusoidal wave with an average period of 2~kpc and maximum amplitude of 60~pc.  The most plausible hypothesis regarding the origin of the Radcliffe wave being the outcome of a large-scale Galactic process of gas accumulation either from a shock front in a spiral arm or from gravitational settling and cooling on the plane of the Milky Way.

\par An analysis of the 3-D distribution of massive stars based on Alma and Gaia DR2 catalogs in the solar neighborhood by \citet{2021MNRAS.tmp..794P} discovered a corrugation pattern,  the Cepheus spur,  extending from Orion-Cygnus arm towards the Perseus arm,  which is likely associated with the Radcliffe wave. Wave-like signatures in the distribution of stars in both density and velocity space have been observed in different regions of the Milky Way where models proposed that the perturbations are caused by the interaction of a satellite galaxy with the disc of the Milky Way \citep[e.g.,][]{2018MNRAS.481.1501B}.  Another theory is that the vertical perturbations can be due to a dynamical coupling between the Galactic bar and the spiral arms \citep[e.g.,][]{2016MNRAS.461.3835M}.  The Gangotri wave is distinct from previously discovered Galactic corrugations in that it is the first velocity coherent dense gas filament exhibiting vertical corrugations.  With the current data,  it is difficult to pinpoint the exact mechanism responsible for the origin of the Gangotri wave. 



\section{Conclusions}

We have identified a new velocity coherent filamentary structure in the fourth Galactic quadrant at $325^\circ<l<355^\circ$, which we call the Gangotri wave.  The structure has a spatial extent of at least 2.0~kpc with a high aspect ratio (60:1) and is morphologically similar to a sinusoidal wave.  A lower limit to the total mass of Gangotri wave is estimated as $8.7\times10^6$ M$_\odot$.  The part of the velocity of the structure is consistent with that of the Norma arm towards one end,  whereas it corresponds to the velocity profile of the 3-kpc arm towards the other end.  The overall velocity structure is consistent with that expected for spiral sub-arm or an inter-arm feather,  similar to those observed in external spiral galaxies.  Density waves,  MHD instabilities,  interaction of the Galactic bar with the spiral arms as well as interaction of external dwarf galaxies with the Milky Way in the past have been suggested as likely scenarios responsible for the formation of such an enigmatic structure.  With the present data,  however,  we cannot conclusively identify its physical origin.    

\section{Acknowledgement}
The authors thank the referee for the comments and suggestions, that improved the quality of the paper.  We thank Zhi Li for providing the spiral arm data based on ($l, v$) model of \citep{2016ApJ...823...77R}. This research made use of Montage which is funded by the National Science Foundation under Grant Number ACI-1440620, and was previously funded by the National Aeronautics and Space Administration's Earth Science Technology Office,  Computation Technologies Project,  under Cooperative Agreement Number NCC5-626 between NASA and the California Institute of Technology. VVS acknowledges support from the Alexander von Humboldt Foundation.

\bibliography{ref}

\begin{thebibliography}{}
\expandafter\ifx\csname natexlab\endcsname\relax\def\natexlab#1{#1}\fi
\providecommand{\url}[1]{\href{#1}{#1}}
\providecommand{\dodoi}[1]{doi:~\href{http://doi.org/#1}{\nolinkurl{#1}}}
\providecommand{\doeprint}[1]{\href{http://ascl.net/#1}{\nolinkurl{http://ascl.net/#1}}}
\providecommand{\doarXiv}[1]{\href{https://arxiv.org/abs/#1}{\nolinkurl{https://arxiv.org/abs/#1}}}

\bibitem[{{Abreu-Vicente} {et~al.}(2016){Abreu-Vicente}, {Ragan},
  {Kainulainen}, {Henning}, {Beuther}, \& {Johnston}}]{2016A&A...590A.131A}
{Abreu-Vicente}, J., {Ragan}, S., {Kainulainen}, J., {et~al.} 2016, \aap, 590,
  A131, \dodoi{10.1051/0004-6361/201527674}

\bibitem[{{Alfaro} {et~al.}(1992){Alfaro}, {Cabrera-Cano}, \&
  {Delgado}}]{1992ApJ...399..576A}
{Alfaro}, E.~J., {Cabrera-Cano}, J., \& {Delgado}, A.~J. 1992, \apj, 399, 576,
  \dodoi{10.1086/171949}

\bibitem[{{Alves} {et~al.}(2020){Alves}, {Zucker}, {Goodman}, {Speagle},
  {Meingast}, {Robitaille}, {Finkbeiner}, {Schlafly}, \&
  {Green}}]{2020Natur.578..237A}
{Alves}, J., {Zucker}, C., {Goodman}, A.~A., {et~al.} 2020, \nat, 578, 237,
  \dodoi{10.1038/s41586-019-1874-z}

\bibitem[{{Barnes} {et~al.}(2015){Barnes}, {Muller}, {Indermuehle},
  {O'Dougherty}, {Lowe}, {Cunningham}, {Hernandez}, \&
  {Fuller}}]{2015ApJ...812....6B}
{Barnes}, P.~J., {Muller}, E., {Indermuehle}, B., {et~al.} 2015, \apj, 812, 6,
  \dodoi{10.1088/0004-637X/812/1/6}

\bibitem[{{Berriman} {et~al.}(2003){Berriman}, {Good}, {Curkendall}, {Jacob},
  {Katz}, {Prince}, \& {Williams}}]{2003ASPC..295..343B}
{Berriman}, G.~B., {Good}, J.~C., {Curkendall}, D.~W., {et~al.} 2003, in
  Astronomical Society of the Pacific Conference Series, Vol. 295, Astronomical
  Data Analysis Software and Systems XII, ed. H.~E. {Payne}, R.~I.
  {Jedrzejewski}, \& R.~N. {Hook}, 343

\bibitem[{{Binney} \& {Sch{\"o}nrich}(2018)}]{2018MNRAS.481.1501B}
{Binney}, J., \& {Sch{\"o}nrich}, R. 2018, \mnras, 481, 1501,
  \dodoi{10.1093/mnras/sty2378}

\bibitem[{{Brand} \& {Blitz}(1993)}]{1993A&A...275...67B}
{Brand}, J., \& {Blitz}, L. 1993, \aap, 275, 67

\bibitem[{{Carlhoff} {et~al.}(2013){Carlhoff}, {Nguyen Luong}, {Schilke},
  {Motte}, {Schneider}, {Beuther}, {Bontemps}, {Heitsch}, {Hill}, {Kramer},
  {Ossenkopf}, {Schuller}, {Simon}, \& {Wyrowski}}]{2013A&A...560A..24C}
{Carlhoff}, P., {Nguyen Luong}, Q., {Schilke}, P., {et~al.} 2013, \aap, 560,
  A24, \dodoi{10.1051/0004-6361/201321592}

\bibitem[{{Colombo} {et~al.}(2014){Colombo}, {Hughes}, {Schinnerer}, {Meidt},
  {Leroy}, {Pety}, {Dobbs}, {Garc{\'\i}a-Burillo}, {Dumas}, {Thompson},
  {Schuster}, \& {Kramer}}]{2014ApJ...784....3C}
{Colombo}, D., {Hughes}, A., {Schinnerer}, E., {et~al.} 2014, \apj, 784, 3,
  \dodoi{10.1088/0004-637X/784/1/3}

\bibitem[{{Dame} {et~al.}(2001){Dame}, {Hartmann}, \&
  {Thaddeus}}]{2001ApJ...547..792D}
{Dame}, T.~M., {Hartmann}, D., \& {Thaddeus}, P. 2001, \apj, 547, 792,
  \dodoi{10.1086/318388}

\bibitem[{{Duarte-Cabral} \& {Dobbs}(2017)}]{2017MNRAS.470.4261D}
{Duarte-Cabral}, A., \& {Dobbs}, C.~L. 2017, \mnras, 470, 4261,
  \dodoi{10.1093/mnras/stx1524}

\bibitem[{{Duarte-Cabral} {et~al.}(2021){Duarte-Cabral}, {Colombo}, {Urquhart},
  {Ginsburg}, {Russeil}, {Schuller}, {Anderson}, {Barnes}, {Beltr{\'a}n},
  {Beuther}, {Bontemps}, {Bronfman}, {Csengeri}, {Dobbs}, {Eden}, {Giannetti},
  {Kauffmann}, {Mattern}, {Medina}, {Menten}, {Lee}, {Pettitt}, {Riener},
  {Rigby}, {Traficante}, {Veena}, {Wienen}, {Wyrowski}, {Agurto}, {Azagra},
  {Cesaroni}, {Finger}, {Gonzalez}, {Henning}, {Hernandez}, {Kainulainen},
  {Leurini}, {Lopez}, {Mac-Auliffe}, {Mazumdar}, {Molinari}, {Motte}, {Muller},
  {Nguyen-Luong}, {Parra}, {Perez-Beaupuits}, {Montenegro-Montes}, {Moore},
  {Ragan}, {S{\'a}nchez-Monge}, {Sanna}, {Schilke}, {Schisano}, {Schneider},
  {Suri}, {Testi}, {Torstensson}, {Venegas}, {Wang}, \&
  {Zavagno}}]{2021MNRAS.500.3027D}
{Duarte-Cabral}, A., {Colombo}, D., {Urquhart}, J.~S., {et~al.} 2021, \mnras,
  500, 3027, \dodoi{10.1093/mnras/staa2480}

\bibitem[{{Elia} {et~al.}(2017){Elia}, {Molinari}, {Schisano}, {Pestalozzi},
  {Pezzuto}, {Merello}, {Noriega-Crespo}, {Moore}, {Russeil}, {Mottram},
  {Paladini}, {Strafella}, {Benedettini}, {Bernard}, {Di Giorgio}, {Eden},
  {Fukui}, {Plume}, {Bally}, {Martin}, {Ragan}, {Jaffa}, {Motte}, {Olmi},
  {Schneider}, {Testi}, {Wyrowski}, {Zavagno}, {Calzoletti}, {Faustini},
  {Natoli}, {Palmeirim}, {Piacentini}, {Piazzo}, {Pilbratt}, {Polychroni},
  {Baldeschi}, {Beltr{\'a}n}, {Billot}, {Cambr{\'e}sy}, {Cesaroni},
  {Garc{\'\i}a-Lario}, {Hoare}, {Huang}, {Joncas}, {Liu}, {Maiolo}, {Marsh},
  {Maruccia}, {M{\`e}ge}, {Peretto}, {Rygl}, {Schilke}, {Thompson},
  {Traficante}, {Umana}, {Veneziani}, {Ward-Thompson}, {Whitworth}, {Arab},
  {Bandieramonte}, {Becciani}, {Brescia}, {Buemi}, {Bufano}, {Butora},
  {Cavuoti}, {Costa}, {Fiorellino}, {Hajnal}, {Hayakawa}, {Kacsuk}, {Leto}, {Li
  Causi}, {Marchili}, {Martinavarro-Armengol}, {Mercurio}, {Molinaro},
  {Riccio}, {Sano}, {Sciacca}, {Tachihara}, {Torii}, {Trigilio}, {Vitello}, \&
  {Yamamoto}}]{2017MNRAS.471..100E}
{Elia}, D., {Molinari}, S., {Schisano}, E., {et~al.} 2017, \mnras, 471, 100,
  \dodoi{10.1093/mnras/stx1357}

\bibitem[{{Elmegreen}(1985)}]{1985IAUS..106..255E}
{Elmegreen}, D.~M. 1985, in The Milky Way Galaxy, ed. H.~{van Woerden}, R.~J.
  {Allen}, \& W.~B. {Burton}, Vol. 106, 255--270

\bibitem[{{Fux}(1999)}]{1999A&A...345..787F}
{Fux}, R. 1999, \aap, 345, 787.
\newblock \doarXiv{astro-ph/9903154}

\bibitem[{{Gaia Collaboration} {et~al.}(2018){Gaia Collaboration}, {Katz},
  {Antoja}, {Romero-G{\'o}mez}, {Drimmel}, {Reyl{\'e}}, {Seabroke}, {Soubiran},
  {Babusiaux}, {Di Matteo}, {Figueras}, {Poggio}, {Robin}, {Evans}, {Brown},
  {Vallenari}, {Prusti}, {de Bruijne}, {Bailer-Jones}, {Biermann}, {Eyer},
  {Jansen}, {Jordi}, {Klioner}, {Lammers}, {Lindegren}, {Luri}, {Mignard},
  {Panem}, {Pourbaix}, {Randich}, {Sartoretti}, {Siddiqui}, {van Leeuwen},
  {Walton}, {Arenou}, {Bastian}, {Cropper}, {Lattanzi}, {Bakker}, {Cacciari},
  {Casta n}, {Chaoul}, {Cheek}, {De Angeli}, {Fabricius}, {Guerra}, {Holl},
  {Masana}, {Messineo}, {Mowlavi}, {Nienartowicz}, {Panuzzo}, {Portell},
  {Riello}, {Tanga}, {Th{\'e}venin}, {Gracia-Abril}, {Comoretto},
  {Garcia-Reinaldos}, {Teyssier}, {Altmann}, {Andrae}, {Audard},
  {Bellas-Velidis}, {Benson}, {Berthier}, {Blomme}, {Burgess}, {Busso},
  {Carry}, {Cellino}, {Clementini}, {Clotet}, {Creevey}, {Davidson}, {De
  Ridder}, {Delchambre}, {Dell'Oro}, {Ducourant},
  {Fern{\'a}ndez-Hern{\'a}ndez}, {Fouesneau}, {Fr{\'e}mat}, {Galluccio},
  {Garc{\'\i}a-Torres}, {Gonz{\'a}lez-N{\'u}{\~n}ez}, {Gonz{\'a}lez-Vidal},
  {Gosset}, {Guy}, {Halbwachs}, {Hambly}, {Harrison}, {Hern{\'a}ndez},
  {Hestroffer}, {Hodgkin}, {Hutton}, {Jasniewicz}, {Jean-Antoine-Piccolo},
  {Jordan}, {Korn}, {Krone-Martins}, {Lanzafame}, {Lebzelter}, {L{\"o}ffler},
  {Manteiga}, {Marrese}, {Mart{\'\i}n-Fleitas}, {Moitinho}, {Mora}, {Muinonen},
  {Osinde}, {Pancino}, {Pauwels}, {Petit}, {Recio-Blanco}, {Richards},
  {Rimoldini}, {Sarro}, {Siopis}, {Smith}, {Sozzetti}, {S{\"u}veges}, {Torra},
  {van Reeven}, {Abbas}, {Abreu Aramburu}, {Accart}, {Aerts}, {Altavilla},
  {{\'A}lvarez}, {Alvarez}, {Alves}, {Anderson}, {Andrei}, {Anglada Varela},
  {Antiche}, {Arcay}, {Astraatmadja}, {Bach}, {Baker},
  {Balaguer-N{\'u}{\~n}ez}, {Balm}, {Barache}, {Barata}, {Barbato}, {Barblan},
  {Barklem}, {Barrado}, {Barros}, {Barstow}, {Bartholom{\'e} Mu{\~n}oz},
  {Bassilana}, {Becciani}, {Bellazzini}, {Berihuete}, {Bertone}, {Bianchi},
  {Bienaym{\'e}}, {Blanco-Cuaresma}, {Boch}, {Boeche}, {Bombrun}, {Borrachero},
  {Bossini}, {Bouquillon}, {Bourda}, {Bragaglia}, {Bramante}, {Breddels},
  {Bressan}, {Brouillet}, {Br{\"u}semeister}, {Brugaletta}, {Bucciarelli},
  {Burlacu}, {Busonero}, {Butkevich}, {Buzzi}, {Caffau}, {Cancelliere},
  {Cannizzaro}, {Cantat-Gaudin}, {Carballo}, {Carlucci}, {Carrasco},
  {Casamiquela}, {Castellani}, {Castro-Ginard}, {Charlot}, {Chemin},
  {Chiavassa}, {Cocozza}, {Costigan}, {Cowell}, {Crifo}, {Crosta}, {Crowley},
  {Cuypers}, {Dafonte}, {Damerdji}, {Dapergolas}, {David}, {David}, {de
  Laverny}, {De Luise}, {De March}, {de Souza}, {de Torres}, {Debosscher}, {del
  Pozo}, {Delbo}, {Delgado}, {Delgado}, {Diakite}, {Diener}, {Distefano},
  {Dolding}, {Drazinos}, {Dur{\'a}n}, {Edvardsson}, {Enke}, {Eriksson},
  {Esquej}, {Eynard Bontemps}, {Fabre}, {Fabrizio}, {Faigler}, {Falc a},
  {Farr{\`a}s Casas}, {Federici}, {Fedorets}, {Fernique}, {Filippi},
  {Findeisen}, {Fonti}, {Fraile}, {Fraser}, {Fr{\'e}zouls}, {Gai}, {Galleti},
  {Garabato}, {Garc{\'\i}a-Sedano}, {Garofalo}, {Garralda}, {Gavel}, {Gavras},
  {Gerssen}, {Geyer}, {Giacobbe}, {Gilmore}, {Girona}, {Giuffrida}, {Glass},
  {Gomes}, {Granvik}, {Gueguen}, {Guerrier}, {Guiraud}, {Guti{\'e}}, {Haigron},
  {Hatzidimitriou}, {Hauser}, {Haywood}, {Heiter}, {Helmi}, {Heu}, {Hilger},
  {Hobbs}, {Hofmann}, {Holland}, {Huckle}, {Hypki}, {Icardi}, {Jan{\ss}en},
  {Jevardat de Fombelle}, {Jonker}, {Juh{\'a}sz}, {Julbe}, {Karampelas},
  {Kewley}, {Klar}, {Kochoska}, {Kohley}, {Kolenberg}, {Kontizas}, {Kontizas},
  {Koposov}, {Kordopatis}, {Kostrzewa-Rutkowska}, {Koubsky}, {Lambert},
  {Lanza}, {Lasne}, {Lavigne}, {Le Fustec}, {Le Poncin-Lafitte}, {Lebreton},
  {Leccia}, {Leclerc}, {Lecoeur-Taibi}, {Lenhardt}, {Leroux}, {Liao}, {Licata},
  {Lindstr{\o}m}, {Lister}, {Livanou}, {Lobel}, {L{\'o}pez}, {Managau}, {Mann},
  {Mantelet}, {Marchal}, {Marchant}, {Marconi}, {Marinoni}, {Marschalk{\'o}},
  {Marshall}, {Martino}, {Marton}, {Mary}, {Massari}, {Matijevi{\v{c}}},
  {Mazeh}, {McMillan}, {Messina}, {Michalik}, {Millar}, {Molina}, {Molinaro},
  {Moln{\'a}r}, {Montegriffo}, {Mor}, {Morbidelli}, {Morel}, {Morris},
  {Mulone}, {Muraveva}, {Musella}, {Nelemans}, {Nicastro}, {Noval},
  {O'Mullane}, {Ord{\'e}novic}, {Ord{\'o}{\~n}ez-Blanco}, {Osborne}, {Pagani},
  {Pagano}, {Pailler}, {Palacin}, {Palaversa}, {Panahi}, {Pawlak},
  {Piersimoni}, {Pineau}, {Plachy}, {Plum}, {Poujoulet}, {Pr{\v{s}}a},
  {Pulone}, {Racero}, {Ragaini}, {Rambaux}, {Ramos-Lerate}, {Regibo}, {Riclet},
  {Ripepi}, {Riva}, {Rivard}, {Rixon}, {Roegiers}, {Roelens}, {Rowell},
  {Royer}, {Ruiz-Dern}, {Sadowski}, {Sagrist{\`a} Sell{\'e}s}, {Sahlmann},
  {Salgado}, {Salguero}, {Sanna}, {Santana-Ros}, {Sarasso}, {Savietto},
  {Schultheis}, {Sciacca}, {Segol}, {Segovia}, {S{\'e}gransan}, {Shih},
  {Siltala}, {Silva}, {Smart}, {Smith}, {Solano}, {Solitro}, {Sordo}, {Soria
  Nieto}, {Souchay}, {Spagna}, {Spoto}, {Stampa}, {Steele},
  {Steidelm{\"u}ller}, {Stephenson}, {Stoev}, {Suess}, {Surdej}, {Szabados},
  {Szegedi-Elek}, {Tapiador}, {Taris}, {Tauran}, {Taylor}, {Teixeira},
  {Terrett}, {Teyssandier}, {Thuillot}, {Titarenko}, {Torra Clotet}, {Turon},
  {Ulla}, {Utrilla}, {Uzzi}, {Vaillant}, {Valentini}, {Valette}, {van Elteren},
  {Van Hemelryck}, {van Leeuwen}, {Vaschetto}, {Vecchiato}, {Veljanoski},
  {Viala}, {Vicente}, {Vogt}, {von Essen}, {Voss}, {Votruba}, {Voutsinas},
  {Walmsley}, {Weiler}, {Wertz}, {Wevers}, {Wyrzykowski}, {Yoldas},
  {{\v{Z}}erjal}, {Ziaeepour}, {Zorec}, {Zschocke}, {Zucker}, {Zurbach}, \&
  {Zwitter}}]{2018A&A...616A..11G}
{Gaia Collaboration}, {Katz}, D., {Antoja}, T., {et~al.} 2018, \aap, 616, A11,
  \dodoi{10.1051/0004-6361/201832865}

\bibitem[{{Georgelin} \& {Georgelin}(1976)}]{1976A&A....49...57G}
{Georgelin}, Y.~M., \& {Georgelin}, Y.~P. 1976, \aap, 49, 57

\bibitem[{{Goodman} {et~al.}(2014){Goodman}, {Alves}, {Beaumont}, {Benjamin},
  {Borkin}, {Burkert}, {Dame}, {Jackson}, {Kauffmann}, {Robitaille}, \&
  {Smith}}]{2014ApJ...797...53G}
{Goodman}, A.~A., {Alves}, J., {Beaumont}, C.~N., {et~al.} 2014, \apj, 797, 53,
  \dodoi{10.1088/0004-637X/797/1/53}

\bibitem[{{Henshaw} {et~al.}(2020){Henshaw}, {Kruijssen}, {Longmore}, {Riener},
  {Leroy}, {Rosolowsky}, {Ginsburg}, {Battersby}, {Chevance}, {Meidt},
  {Glover}, {Hughes}, {Kainulainen}, {Klessen}, {Schinnerer}, {Schruba},
  {Beuther}, {Bigiel}, {Blanc}, {Emsellem}, {Henning}, {Herrera}, {Koch},
  {Pety}, {Ragan}, \& {Sun}}]{2020NatAs...4.1064H}
{Henshaw}, J.~D., {Kruijssen}, J.~M.~D., {Longmore}, S.~N., {et~al.} 2020,
  Nature Astronomy, 4, 1064, \dodoi{10.1038/s41550-020-1126-z}

\bibitem[{{Jackson} {et~al.}(2010){Jackson}, {Finn}, {Chambers}, {Rathborne},
  \& {Simon}}]{2010ApJ...719L.185J}
{Jackson}, J.~M., {Finn}, S.~C., {Chambers}, E.~T., {Rathborne}, J.~M., \&
  {Simon}, R. 2010, \apjl, 719, L185, \dodoi{10.1088/2041-8205/719/2/L185}

\bibitem[{{Kim} {et~al.}(2020){Kim}, {Kim}, \&
  {Ostriker}}]{2020ApJ...898...35K}
{Kim}, W.-T., {Kim}, C.-G., \& {Ostriker}, E.~C. 2020, \apj, 898, 35,
  \dodoi{10.3847/1538-4357/ab9b87}

\bibitem[{{Kim} {et~al.}(2014){Kim}, {Kim}, \& {Kim}}]{2014ApJ...789...68K}
{Kim}, W.-T., {Kim}, Y., \& {Kim}, J.-G. 2014, \apj, 789, 68,
  \dodoi{10.1088/0004-637X/789/1/68}

\bibitem[{{Kim} \& {Ostriker}(2002)}]{2002ApJ...570..132K}
{Kim}, W.-T., \& {Ostriker}, E.~C. 2002, \apj, 570, 132, \dodoi{10.1086/339352}

\bibitem[{{Kuhn} {et~al.}(2021){Kuhn}, {Benjamin}, {Zucker}, {Krone-Martins},
  {de Souza}, {Castro-Ginard}, {Ishida}, {Povich}, \&
  {Hillenbrand}}]{2021A&A...651L..10K}
{Kuhn}, M.~A., {Benjamin}, R.~A., {Zucker}, C., {et~al.} 2021, \aap, 651, L10,
  \dodoi{10.1051/0004-6361/202141198}

\bibitem[{{La Vigne} {et~al.}(2006){La Vigne}, {Vogel}, \&
  {Ostriker}}]{2006ApJ...650..818L}
{La Vigne}, M.~A., {Vogel}, S.~N., \& {Ostriker}, E.~C. 2006, \apj, 650, 818,
  \dodoi{10.1086/506589}

\bibitem[{{Lindner} {et~al.}(2015){Lindner}, {Vera-Ciro}, {Murray},
  {Stanimirovi{\'c}}, {Babler}, {Heiles}, {Hennebelle}, {Goss}, \&
  {Dickey}}]{2015AJ....149..138L}
{Lindner}, R.~R., {Vera-Ciro}, C., {Murray}, C.~E., {et~al.} 2015, \aj, 149,
  138, \dodoi{10.1088/0004-6256/149/4/138}

\bibitem[{{Lynds}(1970)}]{1970IAUS...38...26L}
{Lynds}, B.~T. 1970, in The Spiral Structure of our Galaxy, ed. W.~{Becker} \&
  G.~I. {Kontopoulos}, Vol.~38, 26

\bibitem[{{McClure-Griffiths} {et~al.}(2005){McClure-Griffiths}, {Dickey},
  {Gaensler}, {Green}, {Haverkorn}, \& {Strasser}}]{2005ApJS..158..178M}
{McClure-Griffiths}, N.~M., {Dickey}, J.~M., {Gaensler}, B.~M., {et~al.} 2005,
  \apjs, 158, 178, \dodoi{10.1086/430114}

\bibitem[{{Monari} {et~al.}(2016){Monari}, {Famaey}, {Siebert}, {Grand},
  {Kawata}, \& {Boily}}]{2016MNRAS.461.3835M}
{Monari}, G., {Famaey}, B., {Siebert}, A., {et~al.} 2016, \mnras, 461, 3835,
  \dodoi{10.1093/mnras/stw1564}

\bibitem[{{Nakanishi} \& {Sofue}(2006)}]{2006PASJ...58..847N}
{Nakanishi}, H., \& {Sofue}, Y. 2006, \pasj, 58, 847,
  \dodoi{10.1093/pasj/58.5.847}

\bibitem[{{Narayan} {et~al.}(2020){Narayan}, {Dettmar}, \&
  {Saha}}]{2020MNRAS.495.3705N}
{Narayan}, C.~A., {Dettmar}, R.-J., \& {Saha}, K. 2020, \mnras, 495, 3705,
  \dodoi{10.1093/mnras/staa1400}

\bibitem[{{Oort} {et~al.}(1958){Oort}, {Kerr}, \&
  {Westerhout}}]{1958MNRAS.118..379O}
{Oort}, J.~H., {Kerr}, F.~J., \& {Westerhout}, G. 1958, \mnras, 118, 379,
  \dodoi{10.1093/mnras/118.4.379}

\bibitem[{{Pantaleoni Gonz{\'a}lez} {et~al.}(2021){Pantaleoni Gonz{\'a}lez},
  {Ma{\'\i}z Apell{\'a}niz}, {Barb{\'a}}, \& {Reed}}]{2021MNRAS.tmp..794P}
{Pantaleoni Gonz{\'a}lez}, M., {Ma{\'\i}z Apell{\'a}niz}, J., {Barb{\'a}},
  R.~H., \& {Reed}, B.~C. 2021, \mnras, \dodoi{10.1093/mnras/stab688}

\bibitem[{{Piddington}(1973)}]{1973ApJ...179..755P}
{Piddington}, J.~H. 1973, \apj, 179, 755, \dodoi{10.1086/151913}

\bibitem[{{Ragan} {et~al.}(2014){Ragan}, {Henning}, {Tackenberg}, {Beuther},
  {Johnston}, {Kainulainen}, \& {Linz}}]{2014A&A...568A..73R}
{Ragan}, S.~E., {Henning}, T., {Tackenberg}, J., {et~al.} 2014, \aap, 568, A73,
  \dodoi{10.1051/0004-6361/201423401}

\bibitem[{{Reid} {et~al.}(2016){Reid}, {Dame}, {Menten}, \&
  {Brunthaler}}]{2016ApJ...823...77R}
{Reid}, M.~J., {Dame}, T.~M., {Menten}, K.~M., \& {Brunthaler}, A. 2016, \apj,
  823, 77, \dodoi{10.3847/0004-637X/823/2/77}

\bibitem[{{Reid} {et~al.}(2019){Reid}, {Menten}, {Brunthaler}, {Zheng}, {Dame},
  {Xu}, {Li}, {Sakai}, {Wu}, {Immer}, {Zhang}, {Sanna}, {Moscadelli}, {Rygl},
  {Bartkiewicz}, {Hu}, {Quiroga-Nu{\~n}ez}, \& {van
  Langevelde}}]{2019ApJ...885..131R}
{Reid}, M.~J., {Menten}, K.~M., {Brunthaler}, A., {et~al.} 2019, \apj, 885,
  131, \dodoi{10.3847/1538-4357/ab4a11}

\bibitem[{{Riener} {et~al.}(2019){Riener}, {Kainulainen}, {Henshaw}, {Orkisz},
  {Murray}, \& {Beuther}}]{2019A&A...628A..78R}
{Riener}, M., {Kainulainen}, J., {Henshaw}, J.~D., {et~al.} 2019, \aap, 628,
  A78, \dodoi{10.1051/0004-6361/201935519}

\bibitem[{{Roman-Duval} {et~al.}(2009){Roman-Duval}, {Jackson}, {Heyer},
  {Johnson}, {Rathborne}, {Shah}, \& {Simon}}]{2009ApJ...699.1153R}
{Roman-Duval}, J., {Jackson}, J.~M., {Heyer}, M., {et~al.} 2009, \apj, 699,
  1153, \dodoi{10.1088/0004-637X/699/2/1153}

\bibitem[{{Schinnerer} {et~al.}(2017){Schinnerer}, {Meidt}, {Colombo},
  {Chandar}, {Dobbs}, {Garc{\'\i}a-Burillo}, {Hughes}, {Leroy}, {Pety},
  {Querejeta}, {Kramer}, \& {Schuster}}]{2017ApJ...836...62S}
{Schinnerer}, E., {Meidt}, S.~E., {Colombo}, D., {et~al.} 2017, \apj, 836, 62,
  \dodoi{10.3847/1538-4357/836/1/62}

\bibitem[{{Schuller} {et~al.}(2017){Schuller}, {Csengeri}, {Urquhart},
  {Duarte-Cabral}, {Barnes}, {Giannetti}, {Hernandez}, {Leurini}, {Mattern},
  {Medina}, {Agurto}, {Azagra}, {Anderson}, {Beltr{\'a}n}, {Beuther},
  {Bontemps}, {Bronfman}, {Dobbs}, {Dumke}, {Finger}, {Ginsburg}, {Gonzalez},
  {Henning}, {Kauffmann}, {Mac-Auliffe}, {Menten}, {Montenegro-Montes},
  {Moore}, {Muller}, {Parra}, {Perez-Beaupuits}, {Pettitt}, {Russeil},
  {S{\'a}nchez-Monge}, {Schilke}, {Schisano}, {Suri}, {Testi}, {Torstensson},
  {Venegas}, {Wang}, {Wienen}, {Wyrowski}, \& {Zavagno}}]{2017A&A...601A.124S}
{Schuller}, F., {Csengeri}, T., {Urquhart}, J.~S., {et~al.} 2017, \aap, 601,
  A124, \dodoi{10.1051/0004-6361/201628933}

\bibitem[{{Schuller} {et~al.}(2021){Schuller}, {Urquhart}, {Csengeri},
  {Colombo}, {Duarte-Cabral}, {Mattern}, {Ginsburg}, {Pettitt}, {Wyrowski},
  {Anderson}, {Azagra}, {Barnes}, {Beltran}, {Beuther}, {Billington},
  {Bronfman}, {Cesaroni}, {Dobbs}, {Eden}, {Lee}, {Medina}, {Menten}, {Moore},
  {Montenegro-Montes}, {Ragan}, {Rigby}, {Riener}, {Russeil}, {Schisano},
  {Sanchez-Monge}, {Traficante}, {Zavagno}, {Agurto}, {Bontemps}, {Finger},
  {Giannetti}, {Gonzalez}, {Hernandez}, {Henning}, {Kainulainen}, {Kauffmann},
  {Leurini}, {Lopez}, {Mac-Auliffe}, {Mazumdar}, {Molinari}, {Motte}, {Muller},
  {Nguyen-Luong}, {Parra}, {Perez-Beaupuits}, {Schilke}, {Schneider}, {Suri},
  {Testi}, {Torstensson}, {Veena}, {Venegas}, {Wang}, \&
  {Wienen}}]{2021MNRAS.500.3064S}
{Schuller}, F., {Urquhart}, J.~S., {Csengeri}, T., {et~al.} 2021, \mnras, 500,
  3064, \dodoi{10.1093/mnras/staa2369}

\bibitem[{{Shetty} \& {Ostriker}(2006)}]{2006ApJ...647..997S}
{Shetty}, R., \& {Ostriker}, E.~C. 2006, \apj, 647, 997, \dodoi{10.1086/505594}

\bibitem[{{Simpson} {et~al.}(2012){Simpson}, {Povich}, {Kendrew}, {Lintott},
  {Bressert}, {Arvidsson}, {Cyganowski}, {Maddison}, {Schawinski}, {Sherman},
  {Smith}, \& {Wolf-Chase}}]{2012MNRAS.424.2442S}
{Simpson}, R.~J., {Povich}, M.~S., {Kendrew}, S., {et~al.} 2012, \mnras, 424,
  2442, \dodoi{10.1111/j.1365-2966.2012.20770.x}

\bibitem[{{Smith} {et~al.}(2014){Smith}, {Glover}, {Clark}, {Klessen}, \&
  {Springel}}]{2014MNRAS.441.1628S}
{Smith}, R.~J., {Glover}, S. C.~O., {Clark}, P.~C., {Klessen}, R.~S., \&
  {Springel}, V. 2014, \mnras, 441, 1628, \dodoi{10.1093/mnras/stu616}

\bibitem[{{Smith} {et~al.}(2020){Smith}, {Tre{\ss}}, {Sormani}, {Glover},
  {Klessen}, {Clark}, {Izquierdo}, {Duarte-Cabral}, \&
  {Zucker}}]{2020MNRAS.492.1594S}
{Smith}, R.~J., {Tre{\ss}}, R.~G., {Sormani}, M.~C., {et~al.} 2020, \mnras,
  492, 1594, \dodoi{10.1093/mnras/stz3328}

\bibitem[{{Sormani} {et~al.}(2017){Sormani}, {Sobacchi}, {Shore}, {Tre{\ss}},
  \& {Klessen}}]{2017MNRAS.471.2932S}
{Sormani}, M.~C., {Sobacchi}, E., {Shore}, S.~N., {Tre{\ss}}, R.~G., \&
  {Klessen}, R.~S. 2017, \mnras, 471, 2932, \dodoi{10.1093/mnras/stx1678}

\bibitem[{{Sormani} {et~al.}(2018){Sormani}, {Tre{\ss}}, {Ridley}, {Glover},
  {Klessen}, {Binney}, {Magorrian}, \& {Smith}}]{2018MNRAS.475.2383S}
{Sormani}, M.~C., {Tre{\ss}}, R.~G., {Ridley}, M., {et~al.} 2018, \mnras, 475,
  2383, \dodoi{10.1093/mnras/stx3258}

\bibitem[{{Soto} {et~al.}(2019){Soto}, {Barb{\'a}}, {Minniti}, {Kunder},
  {Majaess}, {Nilo-Castell{\'o}n}, {Alonso-Garc{\'\i}a}, {Leone}, {Morelli},
  {Haikala}, {Firpo}, {Lucas}, {Emerson}, {Moni Bidin}, {Geisler}, {Saito},
  {Gurovich}, {Contreras Ramos}, {Rejkuba}, {Barbieri}, {Roman-Lopes},
  {Hempel}, {Alonso}, {Baravalle}, {Borissova}, {Kurtev}, \&
  {Milla}}]{2019MNRAS.488.2650S}
{Soto}, M., {Barb{\'a}}, R., {Minniti}, D., {et~al.} 2019, \mnras, 488, 2650,
  \dodoi{10.1093/mnras/stz1752}

\bibitem[{{Urquhart} {et~al.}(2021){Urquhart}, {Figura}, {Cross}, {Wells},
  {Moore}, {Eden}, {Ragan}, {Pettitt}, {Duarte-Cabral}, {Colombo}, {Schuller},
  {Csengeri}, {Mattern}, {Beuther}, {Menten}, {Wyrowski}, {Anderson}, {Barnes},
  {Beltr{\'a}n}, {Billington}, {Bronfman}, {Giannetti}, {Kainulainen},
  {Kauffmann}, {Lee}, {Leurini}, {Medina}, {Montenegro-Montes}, {Riener},
  {Rigby}, {S{\'a}nchez-Monge}, {Schilke}, {Schisano}, {Traficante}, \&
  {Wienen}}]{2021MNRAS.500.3050U}
{Urquhart}, J.~S., {Figura}, C., {Cross}, J.~R., {et~al.} 2021, \mnras, 500,
  3050, \dodoi{10.1093/mnras/staa2512}

\bibitem[{{Wada} \& {Koda}(2004)}]{2004MNRAS.349..270W}
{Wada}, K., \& {Koda}, J. 2004, \mnras, 349, 270,
  \dodoi{10.1111/j.1365-2966.2004.07484.x}

\bibitem[{{Wang} {et~al.}(2015){Wang}, {Testi}, {Ginsburg}, {Walmsley},
  {Molinari}, \& {Schisano}}]{2015MNRAS.450.4043W}
{Wang}, K., {Testi}, L., {Ginsburg}, A., {et~al.} 2015, \mnras, 450, 4043,
  \dodoi{10.1093/mnras/stv735}

\bibitem[{{Ward-Thompson} {et~al.}(2010){Ward-Thompson}, {Kirk}, {Andr{\'e}},
  {Saraceno}, {Didelon}, {K{\"o}nyves}, {Schneider}, {Abergel}, {Baluteau},
  {Bernard}, {Bontemps}, {Cambr{\'e}sy}, {Cox}, {di Francesco}, {di Giorgio},
  {Griffin}, {Hargrave}, {Huang}, {Li}, {Martin}, {Men'shchikov}, {Minier},
  {Molinari}, {Motte}, {Olofsson}, {Pezzuto}, {Russeil}, {Sauvage},
  {Sibthorpe}, {Spinoglio}, {Testi}, {White}, {Wilson}, {Woodcraft}, \&
  {Zavagno}}]{2010A&A...518L..92W}
{Ward-Thompson}, D., {Kirk}, J.~M., {Andr{\'e}}, P., {et~al.} 2010, \aap, 518,
  L92, \dodoi{10.1051/0004-6361/201014618}

\bibitem[{{Weaver}(1970)}]{1970IAUS...39...22W}
{Weaver}, H.~F. 1970, in Interstellar Gas Dynamics, ed. H.~J. {Habing},
  Vol.~39, 22

\bibitem[{{Zucker} {et~al.}(2018){Zucker}, {Battersby}, \&
  {Goodman}}]{2018ApJ...864..153Z}
{Zucker}, C., {Battersby}, C., \& {Goodman}, A. 2018, \apj, 864, 153,
  \dodoi{10.3847/1538-4357/aacc66}

\end{thebibliography}
\bibliographystyle{aasjournal}

\appendix

\begin{figure}[h!]
\centering
\hspace*{-1 cm}
\includegraphics[scale=0.47]{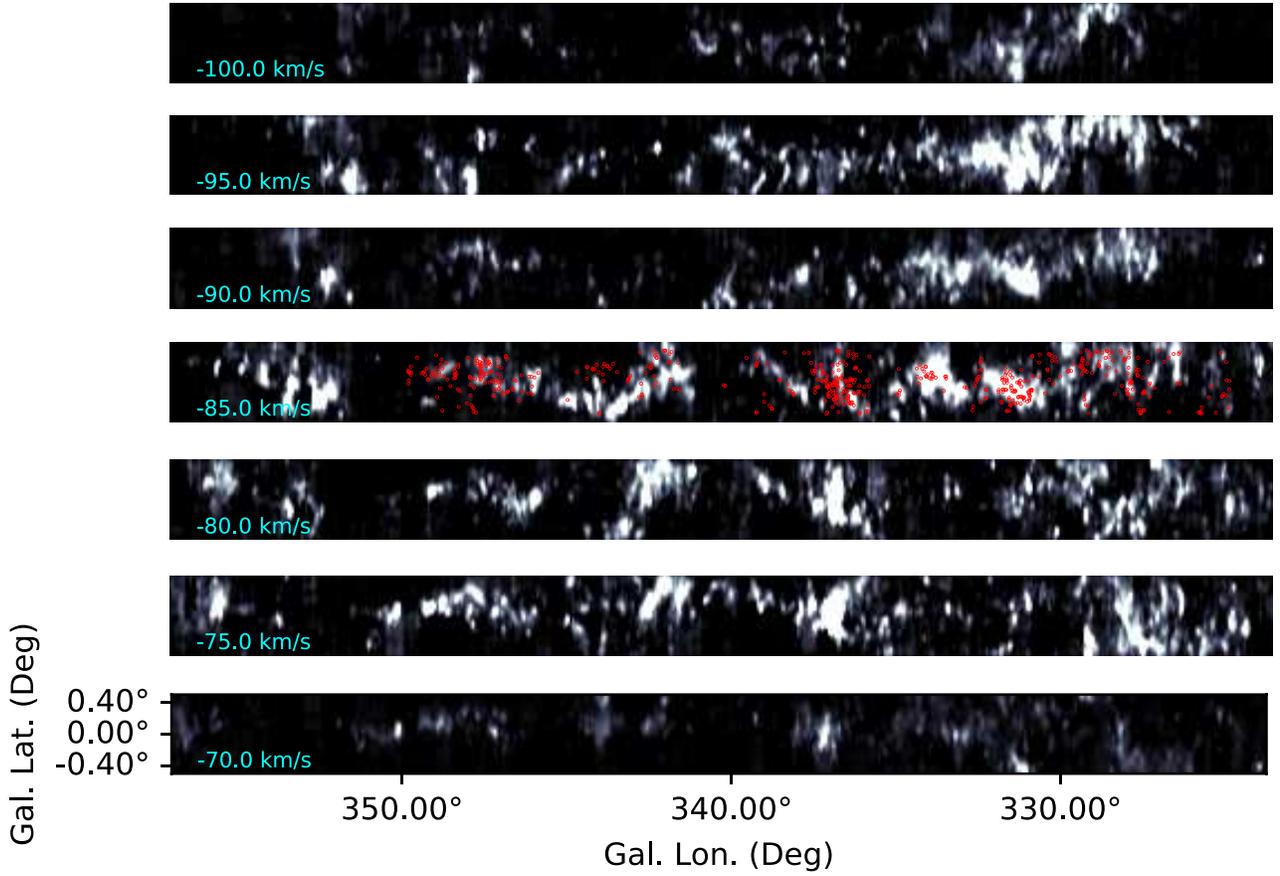} 
\caption{$^{13}$CO channel maps of the Gangotri wave in the velocity range V=[$-100$~km/s, $-70$~km/s].  Overlaid on the V=-85 km/s channel map are the Hi-GAL sources with reliable distance estimates (red). }
\label{cmap}
\end{figure}

\begin{figure}[h!]
\centering
\hspace*{-0.5 cm}
\includegraphics[scale=1.05]{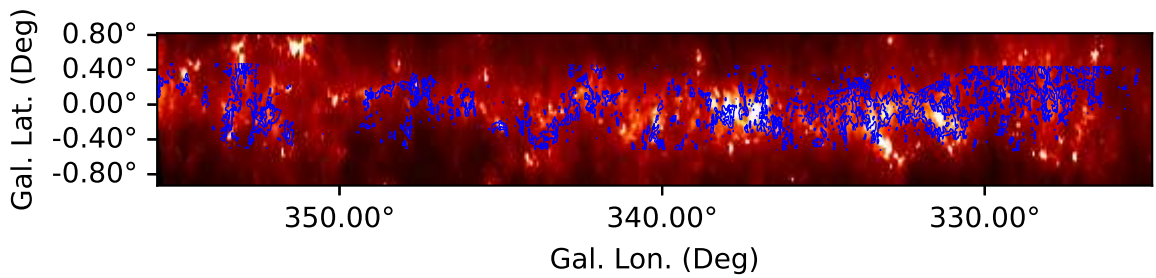} 
\caption{Dust continuum map from $Herschel$ at 500~$\mu$m overlaid with $^{13}$CO integrated intensity contours.  The contour levels are 150~K\,km/s and 450~K\,km/s. The image is stretched along the Y axis for better visualization.}
\label{cmap}
\end{figure}

\end{document}